# Effects of mechanical treatment and exchanged cation on the microporosity of vermiculite

M.C. Jiménez De Haro[a], J.M. Martínez Blanes[a], J. Poyato[a], L.A. Pérez-Maqueda[a],
A. Lerf[b], J.L. Pérez-Rodríguez[a],*

[a]*Instituto de Ciencia de Materiales de Sevilla, CSIC, Universidad de Sevilla, 41092 Sevilla, Spain*
[b]*Walther-Meissner-Institut, Bayerische Akademie der Wissenschaften, D-85748 Garching, Germany*

### Abstract

The influence of the interlayer cations, sonication and grinding on surface area of the vermiculite from Santa Olalla has been studied. For untreated samples the surface area increases with the size of the exchanged cation and only the samples with the largest cations ($NH_4^+$ and $K^+$) show some microporosity. Sonication produces an increase of the surface area but the amount of micropores remains almost unchanged. Grinding leads to a significant increase of the surface area and of the microporosity. In the ground samples the percentage of micropores contributing to the surface area is influenced by the interlayer cation and the extent of $N_2$ penetration by the water kept in the sample at a given degassing temperature. From transmission electron microscopy, thermogravimetry and XRD ($d_{00l}$ diffraction peaks) it can be concluded that grinding produces imperfections in crystal ordering, some alteration in the crystal edges, and a significant release of loosely bound water at temperatures up to 450 °C. These effects are responsible for the high surface area and microporosity measured by the $N_2$ adsorption.

## 1. Introduction

The reduction of the lateral size and particle thickness allows the use of vermiculites in many important applications such as coating, lightweight additive, isolation, etc. Different methods have been proposed for delaminating and reducing the particle size of vermiculites, such as wet or dry grinding. Grinding of vermiculite produces considerable particle size reduction in a first stage, but longer treatments lead to an intense structural degradation of vermiculite with loss of the lamellar shape and progressive amorphization accompanied with formation of hard agglomerates by cold-welding [1]. A method proposed very recently for particle size reduction of clays is sonication. This method has been used for preparing nanometric vermiculite particles [2,3]. Sonication produces delamination in the [00l] direction and breaking of layers in the other crystallographic directions, while the crystalline character is retained [4].

The porosity of clays ranging from micropores to mesopores may originate from: (i) interlayer regions if the layer separation is sufficiently large, (ii) staggered layers, (iii) voids created by the overlapping of stacked layers, and (iv) crevices in the particle surface [5]. The interlayer space in dry clay samples, where the exchangeable cations are located, is strongly restricted. Thus, some workers [6] suggest that the $N_2$ penetration would be nearly impossible because of the excessive energy required to expand the interlayer space and, consequently, the surface area determined is a measure of the external area only. In agreement with this assumption the surface area studies on homoionic vermiculite using nitrogen and carbon dioxide as adsorbates have shown that only very slight penetration occurs between individual platelets [7]. The extent of $N_2$ and $CO_2$ penetration through the edges of the vermiculite platelets is influenced by the coordinated water retained within the sample at a given degassing temperature [7]. Forces between layers are weakened with increasing water content, which permits slightly greater penetration of the adsorbed gases. These results contrast with the behaviour of lower charge phyllosilicates, such as montmorillonite, were it is found that the degree of penetration between layers is quite high and it is governed by the particle size and the interlayer cation [5,7–9]. The effect of mechanical treatments on surface area and microporosity of vermiculites has not been previously performed.

* Corresponding author. Fax: +34-95-4460665.
   E-mail address: jlperez@cica.es (J.L. Pérez-Rodríguez).

In this study we investigate the effect of mechanical treatment and cation type on the clay microporosity of the Santa Olalla vermiculite untreated and mechanically treated (sonicated and ground) and saturated with different cations ($NH_4^+$, $K^+$, $Mg^{2+}$, $Na^+$). The combined data of $d_{00l}$ diffraction measured in a high temperature chamber, thermogravimetry (TG), and $N_2$ vapour adsorption are used for the evaluation of the accessibility of interlayer space to $N_2$.

## 2. Experimental

Vermiculite from Santa Olalla (Huelva, Spain) was used as starting material [2]. The fraction <80 mm was obtained with a knife mill (Retsch ultracentrifuge mill, model 255M-1 equipped with a suitable sieve). Cation exchange was performed with solutions of the corresponding salts $MgCl_2$, NaCl, KCl, and $NH_4CH_3COO$.

The vermiculite sample was sonicated for 40 h with an ultrasonic horn at 20 °C [2]. Grinding experiments were carried out on batches of 10 g each using a vibratory mill (Herzog HSM-100), which works through friction and impact at 1500 rpm.

The adsorption of $N_2$ was determined with a Micromeritic 2200 A Model, Norcross GA. The samples were out-gassed by heating at 175 or 450 °C in vacuum for 8 h. The surface area was determined using the BET method. The micropore volume and the open surface area of the vermiculite were determined by both, the $t$-method and the $a_s$ method [10]. TG experiments were performed in air atmosphere with a Seiko TG/DTA 6300 equipment at a heating rate of 10 K min$^{-1}$ up to 1200 °C. X-ray diffraction measurements were carried out on a Philips diffractometer X'Pert equipped with an Anton Parr high temperature chamber using Cu $K_a$ radiation. Transmission electron microscopy (TEM) measurements were carried out with a Philips CM 200.

## 3. Results and discussion

### 3.1. $N_2$ vapour adsorption

*Untreated vermiculite (<80 mm).* Adsorption isotherms of $N_2$ at liquid $N_2$ temperature on vermiculite saturated with $NH_4^+$, $K^+$, $Na^+$, and $Mg^{2+}$ ions are similar and of type II in shape [10]; as an example the adsorption isotherm of the $NH_4^+$ saturated vermiculite is shown in Fig. 1a.

The total surface areas, as assayed from BET plots, for $NH_4$-, K-, Na-, and Mg- vermiculites are 22.9, 17.8,

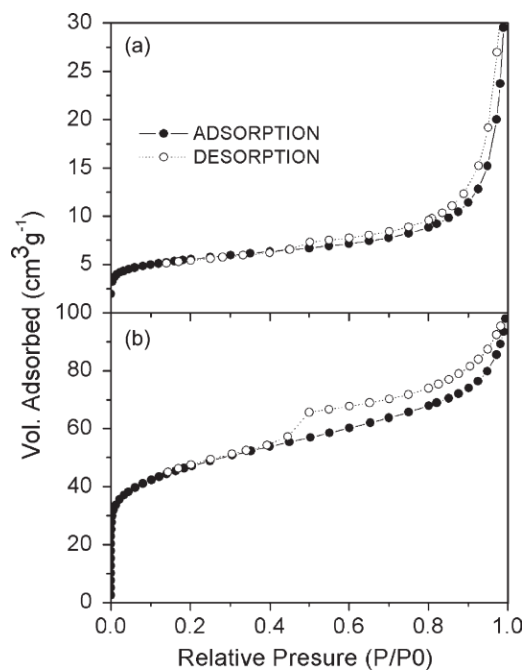

Fig. 1. Adsorption isotherms of (a) untreated vermiculite exchanged with $NH_4^+$ and (b) ground vermiculite and later on exchanged with $NH_4^+$

11.6, and 7.1 m$^2$ g$^{-1}$, respectively. The samples intercalated by $Na^+$ and $Mg^{2+}$ show similar values of total surface area corresponding to external surface area without any microporosity, as indicated by the $t$-plot analysis. On the other hand, $NH_4$- and K-vermiculite show some microporosity (5.19 and 1.9 m$^2$ g$^{-1}$, respectively) that may be due to the larger ionic radio of $NH_4^+$(1.45 Å) and $K^+$(1.33 Å) as compared with those of $Na^+$ (0.98 Å) and $Mg^{2+}$ (0.65 Å). Thus, the larger ionic radii seems to allow a better penetration of $N_2$ in the interlayer space.

*Sonicated vermiculite.* Sonicated vermiculite samples consist of nanometric flakes without changes in the crystallinity of this material [2–4]. The sonicated sample was intercalated with $NH_4^+$, which produces the highest values of total surface and micropore area in the untreated samples. The adsorption isotherm (figure not shown) is very similar to that of the untreated ammonium saturated sample. However, the surface area of the $NH_4^+$ sonicated sample (38.8 m$^2$ g$^{-1}$) is larger than that of the untreated $NH_4^+$ sample (22.9 m$^2$ g$^{-1}$) due to the decrease of particle size caused by sonication. Nevertheless, the micropore area (6.1 m$^2$ g$^{-1}$) remains almost unchanged. These results show that the decrease in particle size of vermiculite produces an increase of surface area while the micropore area does not change.

*Ground vermiculite.* Experiments were performed with a sample ground for 2 min because it showed the highest surface area. In addition, longer treatment times produced significant amorphization. The $N_2$ adsorption isotherms were all very similar for the different interlayer cations, showing again type II adsorption isotherms. As compared with the isotherms of the untreated samples, the isotherms of the ground samples are shifted to higher values of adsorbed volumes and they present some hysteresis. The adsorption isotherm of $NH_4$-vermiculite is shown in Fig. 1b. The total surface areas from BET plots and the micropore areas from *t* and $a_s$ plots for the ground vermiculite and later on saturated with different cations are given in Table 1. We must keep in mind that the values of BET surface area for microporous materials are subject of some errors due to pore filling of the adsorbate [5,10]. Total surface areas, micropore surface areas and micropore volumes experience a strong increase in comparison with untreated and sonicated samples. For ground samples, unlike untreated samples, no correlation is observed between the cation size and the surface area. Nevertheless, for these two series of samples the percentage of surface area contributed by micropores is larger for the vermiculites with larger cations.

Table 1
BET surface area, total pore volume, micropore surface area, micropore volume (*t*-method and $a_s$-method) and percentage of surface area contributed by micropores at 175 and 450 °C degassing temperature of ground vermiculite followed by saturation with different cations

| Cation | Degassing temperature | | | | | | | | | |
|---|---|---|---|---|---|---|---|---|---|---|
| | 175 °C | | | | | 450 °C | | | | |
| | BET surface area ($m^2 g^{-1}$) | Pore volume ($cm^3 g^{-1}$) | Micropore volume *t*-method ($cm^3 g^{-1}$) | Micropore surface area *t*-method ($m^2 g^{-1}$) | Micropore volume $a_s$-method ($cm^3 g^{-1}$) | Surface area contributed by micropore (%) | BET surface area ($m^2 g^{-1}$) | Micropore surface area *t*-method ($m^2 g^{-1}$) | Surface area contributed by micropore (%) |
| $Mg^{2+}$ | 103.8 | 0.097 | 0.010 | 23.70 | 0.009 | 22.2 | 88.1 | 24.0 | 27.2 |
| $Na^+$ | 164.2 | 0.127 | 0.033 | 72.24 | 0.028 | 44.0 | 138.2 | 56.3 | 40.7 |
| $K^+$ | 123.3 | 0.098 | 0.027 | 58.35 | 0.024 | 47.3 | 89.1 | 38.6 | 43.0 |
| $NH_4^+$ | 164.6 | 0.152 | 0.024 | 52.17 | 0.022 | 31.7 | 151.3 | 44.1 | 29.0 |

The only exception to this rule is in the case of $NH_4^+$ ground sample. This unexpected behaviour could be explained by the behaviour during degassing. During the degassing treatment at 175 °C small quantity of ammonium from the interlayer space is released, as previously observed by evolved gas analysis measurements [11]. The release of such ammonium might reduce the microporosity contribution to the surface area.

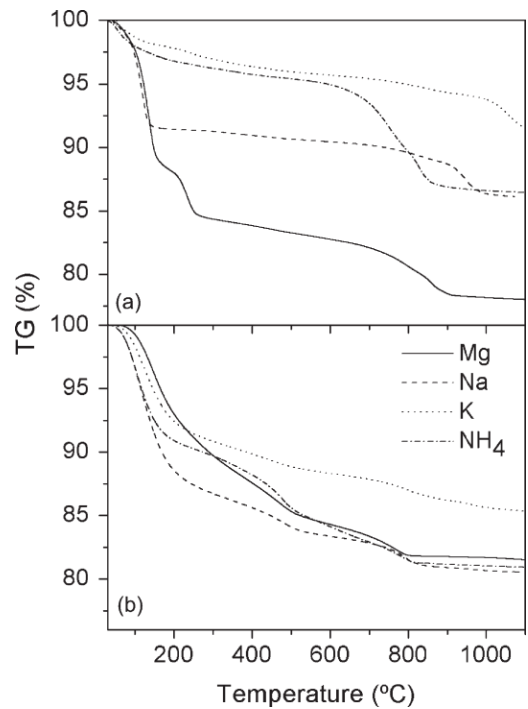

Fig. 2. TG curves of vermiculite saturated with different cations, (a) untreated and (b) ground.

Table 2
Weight loss of ground vermiculite exchanged with different cations and degassed at 175 and 450 8C, and difference in weight loss, surface area and micropore surface area between samples degasses at this two temperatures.

| Cation | Weight loss 175 8C (%) | Weight loss 450 8C (%) | Difference in weight loss 450–175 8C (%) | Difference in surface area 450–175 8C ($m^2 g^{-1}$) | Difference in micropore surface area 450–175 8C ($m^2 g^{-1}$) |
|---|---|---|---|---|---|
| $Mg^{2+}$ | 12.5 | 17.7 | 5.2 | 15.7 | 0 |
| $Na^+$ | 8.8 | 16.4 | 7.6 | 26.0 | 15.9 |
| $K^+$ | 4.8 | 12.8 | 8.0 | 34.2 | 19.7 |
| $NH_4^+$ | 9.0 | 14.7 | 5.7 | 13.3 | 8.1 |

*3.2. Water effect on porosity*

In Section 3.1. it has been shown that for vermiculite both the exchangeable cations and the mechanical treatment produce a significant effect on the porosity of the material. Besides these effects it has been claimed in literature [7,8] that the coordinated water retained within the sample at a given degassing temperature may also play a central role in the adsorption processes. In this section we will study the combined effect of water, grinding, and the interlayer cation on the porosity of vermiculite.

Fig. 2 shows the TG curves for untreated and ground vermiculite samples exchanged with different cations. Untreated and ground materials have very different thermal behaviour [12]. For the untreated material the dehydration and dehydroxylation steps are well resolved. Hydration water is totally released at 220 8C when heated under dynamic conditions as those used in the TG experiment. On the other hand, for the ground materials these steps are not so well resolved and they present continuous weight losses in the entire temperature range from 100 to 450 8C.

In order to study the influence of the remaining water in the sample on the surface area and microporosity of the ground vermiculite, another series of vapour adsorption experiments were performed under identical conditions as those in Fig. 1, but degassing the samples at 450 8C instead of 175 8C to remove all the possible hydration or loosely bound water. The values of BET surface area, micropore surface area, and surface area contributed by micropores for the sample degassed at 450 8C are also included in Table 1 for comparison with those degassed at 175 8C. The BET surface area and micropore areas show a decrease in comparison with the sample degassed at lower temperature. However, the surface area contributed by micropores follows the same order. The smallest decrease on surface and micropore area is shown by vermiculite intercalated by $Mg^{2+}$ and $NH_4^+$ followed by vermiculite intercalated by $K^+$ and $Na^+$. These data are correlated with the difference in weight loss between samples heated at 175 and 450 8C (Table 2). Thus, the samples retaining a larger amount of water when heated at 175 8C (Na- and K-vermiculite) suffer a more significant decrease in the surface area and micropore surface when degassed at 450 8C. These results show that the extent of nitrogen penetration in the vermiculite is influenced by water retained within the sample at a given degassing temperature. Forces between layers and particles are weakened with increasing water content permitting a slightly larger penetration of the adsorbate.

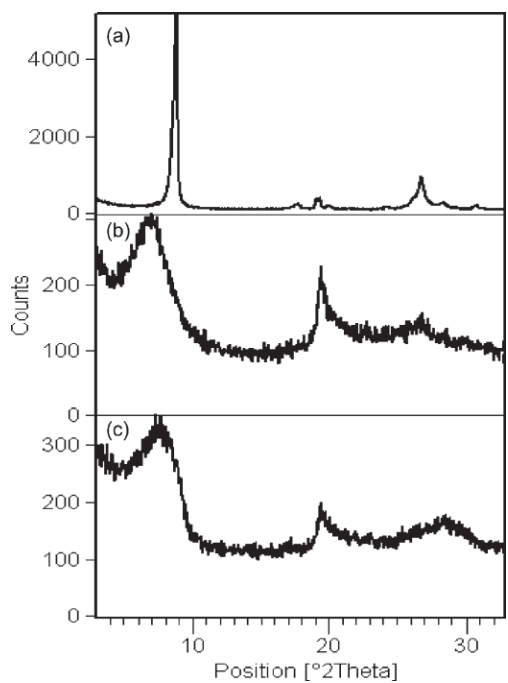

Fig. 3. X-ray diffraction patterns of Mg-vermiculite: (a) untreated and degassed at 175 8C, (b) ground and degassed at 175 8C and (c) ground and degassed at 450 8C.

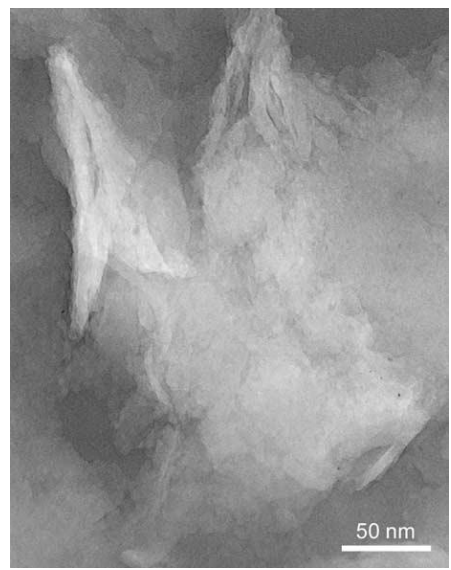

Fig. 4. TEM of ground Mg-vermiculite.

### 3.3. X-ray diffraction and TEM study

The $d_{00l}$ reflections for dry samples (no water in the interlayer space) combined with vapour adsorption data may be used to evaluate the probable location and the accessibility of clay micropores, especially the relative accessibility of the interlayer space. Fig. 3 shows the XRD pattern of the Mg-vermiculite untreated and degassed at 175 8C, ground and degassed at 175 8C, and ground and degassed at 450 8C. The ground vermiculites exchanged with different cations and heated at 175 8C show a significant broadening of the *00l* peak as compared with untreated and sonicated samples. After heating at 450 8C the peak maximum was shifted to lower spacing and remained much broader for the ground samples. Although there are several reasons to explain the width of the diffraction peak such as instrumental reasons, microdomain size, strain broadening, and imperfection in crystal order, the TEM study reveals for the ground sample an important alteration mainly in the edges of the layers and not an important decrease of particle size (Fig. 4). The random displacement and imperfection in crystal order mainly produced by edge alteration may be responsible for the big width of the diffraction $d_{00l}$. Also the statistical arrangement of the layers and the overlap of quasi-crystalline regions may vary as function of the exchangeable cations size and cause broadening of the $d_{00l}$ reflections. These suggestions agree with the high surface and microporosity of surface area of the ground sample.

### Acknowledgements

The authors are grateful to the Ministry of Science and Technology of Spain and to the FEDER program of the EC for economic support through grant (MAT2002-03774).